\documentclass[12pt]{article}
\usepackage{graphicx,amssymb,amsfonts,amsmath,epsfig,color}
\usepackage{mathrsfs}
\usepackage{cite}
\usepackage{url}
\newcounter{mnotecount}%[section]

\newcommand{\mnotex}[1]%{}
{\protect{\stepcounter{mnotecount}}$^{\mbox{\footnotesize $\bullet$\themnotecount}}$
\marginpar{%\color{red}%
\raggedright\tiny\em
$\!\!\!\!\!\!\,\bullet$\themnotecount: #1} }

\makeatletter
\DeclareSymbolFont{AMSb}{U}{msb}{m}{n}
\DeclareSymbolFontAlphabet{\mathbb}{AMSb}
\renewcommand{\section}{\@startsection{section}{1}{\z@}%
                                    {-7ex \@plus -1ex \@minus -.2ex}%
                                    {2.5ex \@plus.2ex}%
                                    {\normalfont\large\scshape\centering}}
\renewcommand{\subsection}{\@startsection{subsection}{2}{\z@}%
                                       {-5ex \@plus -1ex \@minus -.2ex}%
                                       {1.5ex \@plus.2ex}%
                                       {\normalfont\normalsize\scshape}}
\renewcommand{\subsubsection}{\@startsection{subsubsection}{3}{\z@}%
                                       {-5ex \@plus -1ex \@minus -.2ex}%
                                       {1.5ex \@plus.2ex}%
                                       {\normalfont\normalsize\scshape}}

\renewcommand\@seccntformat[1]{\ignorespaces\csname #1name\endcsname\space
                               \csname the#1\endcsname.\quad}   % Extra period and name added
\newdimen\captionmargin
\newdimen\captionindent
\newdimen\captionwidth
\newcommand{\captionfont}{\slshape}
\newcommand\@captionlabel[1]{\textsc{#1:}\space}
\long\def\@makecaption#1#2{%
  \vskip\abovecaptionskip
  \captionwidth\hsize
  \advance\captionwidth -2\captionmargin
  \sbox\@tempboxa{\@captionlabel{#1}\captionfont #2}%
  \ifdim \wd\@tempboxa >\captionwidth
    \ifdim\captionindent>\z@
      \advance\captionwidth -\captionindent
      \hskip\captionindent
    \fi
    \hskip\captionmargin
    \parbox[t]{\captionwidth}{\leavevmode\hskip-\captionindent
      \@captionlabel{#1}\captionfont #2}%
  \else
    \global \@minipagefalse
    \hb@xt@\hsize{\hfil\box\@tempboxa\hfil}%
  \fi
  \vskip\belowcaptionskip}
\def\eqnarray{%
   \stepcounter{equation}%
   \def\@currentlabel{\p@equation\theequation}%
   \global\@eqnswtrue
   \m@th
   \global\@eqcnt\z@
   \tabskip\@centering
   \let\\\@eqncr
   $$\everycr{}\halign to\displaywidth\bgroup
       \hskip\@centering$\displaystyle\tabskip\z@skip{##}$\@eqnsel
      &\global\@eqcnt\@ne$\;\hfil{##}$\hfil
      &\global\@eqcnt\tw@$\;\displaystyle{##}$\hfil\tabskip\@centering
      &\global\@eqcnt\thr@@ \hb@xt@\z@\bgroup\hss##\egroup
         \tabskip\z@skip
      \cr}
\ifcase \@ptsize
\or
\or
\fi
  \divide\@tempdima\baselineskip
  \@tempcnta=\@tempdima
\makeatother
\begin{document}

\renewcommand{\theequation}{\arabic{section}.\arabic{equation}}
\renewcommand{\thefigure}{\arabic{figure}}
\newcommand{\gapprox}{%
\mathrel{%
\setbox0=\hbox{$>$}\raise0.6ex\copy0\kern-\wd0\lower0.65ex\hbox{$\sim$}}}
\textwidth 165mm \textheight 220mm \topmargin 0pt \oddsidemargin 2mm
\def\ib{{\bar \imath}}
\def\jb{{\bar \jmath}}

\newcommand{\ft}[2]{{\textstyle\frac{#1}{#2}}}
\newcommand{\be}{\begin{equation}}
\newcommand{\ee}{\end{equation}}
\newcommand{\bea}{\begin{eqnarray}}
\newcommand{\eea}{\end{eqnarray}}
\newcommand{\Identity}{{1\!\rm l}}% Unit Matrix
\newcommand{\cx}{\overset{\circ}{x}_2}
\def\CN{$\mathcal{N}$}
\def\CH{$\mathcal{H}$}
\def\hg{\hat{g}}
\newcommand{\bref}[1]{(\ref{#1})}
\def\espai{\;\;\;\;\;\;}
\def\zespai{\;\;\;\;}
\def\avall{\vspace{0.5cm}}
\newtheorem{theorem}{Theorem}
\newtheorem{acknowledgement}{Acknowledgment}
\newtheorem{algorithm}{Algorithm}
\newtheorem{axiom}{Axiom}
\newtheorem{case}{Case}
\newtheorem{claim}{Claim}
\newtheorem{conclusion}{Conclusion}
\newtheorem{condition}{Condition}
\newtheorem{conjecture}{Conjecture}
\newtheorem{corollary}{Corollary}
\newtheorem{criterion}{Criterion}
\newtheorem{defi}{Definition}
\newtheorem{example}{Example}
\newtheorem{exercise}{Exercise}
\newtheorem{lemma}{Lemma}
\newtheorem{notation}{Notation}
\newtheorem{problem}{Problem}
\newtheorem{prop}{Proposition}
\newtheorem{rem}{{\it Remark}}
\newtheorem{solution}{Solution}
\newtheorem{summary}{Summary}
\numberwithin{equation}{section}
\newenvironment{pf}[1][Proof]{\noindent{\it {#1.}} }{\ \rule{0.5em}{0.5em}}
\newenvironment{ex}[1][Example]{\noindent{\it {#1.}}}

\thispagestyle{empty}

%\begin{flushright}\scshape
%\end{flushright}
%\vskip1cm

\begin{center}

{\LARGE\scshape Observational and theoretical aspects of Superspinars \par}
\vskip15mm

\textsc{Ram\'{o}n Torres\footnote{E-mail: ramon.torres-herrera@upc.edu}}
\par\bigskip
{\em
Dept. de F\'{i}sica, Universitat Polit\`{e}cnica de Catalunya, Barcelona, Spain.}\\[.1cm]
%

%\vspace{5mm}

\end{center}

\begin{abstract}
This article delves into the observational signatures and theoretical underpinnings of rotating astrophysical objects, with a particular focus on superspinars —exotic objects characterized by the absence of event horizons due to their high angular momentum.

While solutions within General Relativity (\textit{Kerr superspinars}) predict such objects, their classical forms harbor naked singularities, violate causality, and exhibit problematic repulsive gravitational effects. These characteristics render classical superspinars theoretically objectionable, leading to the consideration of them as physically implausible. On the other hand, the incompatibility between General Relativity and Quantum Mechanics suggests the exploration of alternative models, particularly those in which Quantum Gravity dominates the core and prevents the formation of scalar curvature singularities.

This work demonstrates that superspinars without scalar curvature singularities can avoid all the complications associated with Kerr superspinars. Moreover, from a phenomenological standpoint, it is shown that the silhouettes of these superspinars could be markedly distinct from those of black holes and classical Kerr superspinars. To substantiate these differences, we perform a comprehensive analysis of inner null geodesics and investigate the structure of the Planckian region within superspinars without scalar curvature singularities. Our study reveals that only these superspinars provide the potential for distant observers to directly observe the extremely high curvature regions within their interiors.

\end{abstract}


\setcounter{equation}{0}

\section{Introduction}

Recent observational advancements
are enabling us to test our theoretical predictions about rotating black holes. Notably, one confirmed prediction is that observed black holes exhibit high angular momentum, consistent with our theoretical and numerical models. This confirmation prompts us to consider the existence of another theoretical prediction: \textit{superspinars}. Superspinars are horizonless objects possessing a high angular momentum and that satisfy two conditions: (1) the magnitude of their angular momentum $a$ exceeds a threshold $|a|>a^*$; and (2) if their angular momentum were reduced to $|a|<a^*$, the superspinar would transition into a black hole. The threshold $a^*$ is determined by the parameters defining the superspinar.

Note that, when modeling a superspinar, the defining parameters depend on the specific theoretical framework of gravity employed. Consequently, a spinning object with a given total mass and angular momentum may have different threshold values $a^*$ in different gravitational theories and, thus, could be a black hole in certain theoretical approaches while being considered a superspinar in others.

Archetypal superspinars emerge as solutions within the framework of General Relativity, characterized by angular momentum ($a$), mass ($M$) and threshold $a^*=M$. They exceed this threshold since they satisfy $|a| > M$.
These solutions, known as hyperextreme Kerr solutions or \textit{Kerr superspinars}, face significant theoretical challenges. They are associated with naked inner singularities, violations of the cosmic censorship conjecture, causality issues, and problematic repulsive gravitational effects. Consequently, they are often dismissed as physically unrealistic.

Conventionally, Kerr's solution is not expected to be valid in the Planckian regime. It is postulated that Quantum Gravity effects would modify such objects, potentially eliminating their inner scalar curvature singularities. As one of the goals in this article, we aim to demonstrate that, if these superspinars indeed exist, they would be free from the issues that plague Kerr superspinars.

Other significant issues include the origin and stability of superspinars. Two mechanisms have been proposed by which superspinars might exist. First, they could just be born as primordial superspinars. This possibility has been predicted in both the classical framework (see, for instance, \cite{stu2012}) and Quantum Gravity frameworks (see, for example, \cite{G&Hof}\cite{NJGK2017}). After their creation, collapsing matter might reduce the angular momentum of the superspinars, potentially transforming them into black holes. However, it has been demonstrated that this possibility is limited to cases where sufficient surrounding matter exists; otherwise, no event horizon would form, and they would remain as superspinars \cite{SHT}.

It has been shown through various methods that it is impossible to overspin a classical Kerr black hole to transform it into a classical Kerr superspinar. The third law of black hole thermodynamics asserts that no physical process can convert a black hole with $a^2 < M^2$ into an extremal one with $a^2 = M^2$ in a finite number of steps \cite{BCH}. Moreover, such a process would violate the Weak Cosmic Censorship Hypothesis \cite{WCCC}, which stipulates that the singularity should remain concealed from distant observers.

However, within various approaches to quantum gravity, it may be possible to bypass the third law of black hole thermodynamics by overspinning a quantum-enhanced black hole. Some examples of this phenomenon can be found, for instance, in \cite{Li&Bambi, YZWL, EH2023}. Furthermore, in such cases, the Weak Cosmic Censorship Hypothesis does not apply because the resulting superspinar would be nonsingular (and would lie outside the framework of General Relativity). As a result, this procedure could represent a second possible scenario for the existence of superspinars without scalar curvature singularities, which, from now on, we will denote as \textit{scalar-regular}-superspinars or \emph{SR}-superspinars.

In this article, we explore various phenomenological and theoretical aspects of SR--superspinars in contrast to classical superspinars, building upon existing literature. Previous studies on the phenomenology of classical superspinars can be found, for example, in \cite{H&Maeda, B&F, Vries2000, SS2010, TT24}, while some instances of SR-superspinars have been addressed in the excellent work \cite{KG21}. Our objective is to enhance these investigations by conducting a meticulous analysis of the interior of the superspinars. This will utilize recent advancements in understanding the interiors of rotating astrophysical objects \cite{TorresInt}, a detailed examination of the behaviours of null geodesics, and an assessment of the expected shape of the Planckian regions in SR-superspinars.

Through these efforts, we aim to address several key questions: Do SR-superspinars exhibit the same issues as classical superspinars? How do SR-superspinars appear to distant observers compared to rotating black holes or Kerr superspinars? Given the absence of horizons in superspinars, is it true that distant observers will necessarily see the high-curvature region inside superspinars?

These questions will be addressed as follows: In Section \ref{KRBHs}, we examine the general metric of superspinars with a focus on the conditions needed for the absence of scalar curvature singularities, the possibility of avoiding extensions through their disk, the global spacetime structure and the absence of causality violations in the scalar-regular scenario. Section \ref{ecs} is dedicated to identifying the Planckian region of the SR-superspinars, where curvature, effective densities and effective pressures reach Planckian values. Section \ref{ngd} provides a detailed analysis of null geodesics that reach and/or traverse the disk or ring.
The shape of superspinars as perceived by distant observers is investigated in Section \ref{secPheno}. This includes an examination of the images produced by their unstable circular orbits and the portrayal of the Planckian region as observed from a distance, both in classical and scalar-regular cases. Section \ref{SCOs} is devoted to stable circular orbits and their potentially associated instabilities. Finally, Section \ref{conclu} summarizes the findings of this article.

\section{Metric of Scalar-Regular-Superspinars}
\label{KRBHs}

Presently, the absence of a fully developed and reliable quantum theory of gravity poses a significant challenge in accurately describing quantum black holes and superspinars. Therefore, it is crucial to explore phenomenological approaches to assess potential models of rotating objects without scalar curvature singularities and their implications, including the potential for observable astrophysical predictions.

There are numerous proposals for rotating black holes spacetimes without scalar curvature singularities with their corresponding metrics in the literature (see, for example,
\cite{A-A,B&M,FLMV,LGS,Maeda,MFL,eye,D&G,Ghosh,Tosh,R&T,TorresExt,BMR,G&H,Buri,C&M,S&S}).
These proposals share many common properties, including a critical feature akin to Kerr's solution: if the angular momentum exceeds a specific threshold, the model lacks an event horizon and therefore does not represent a rotating \emph{black hole}, but rather what we term a \textit{SR-superspinar}.

The general metric for such rotating models was discovered by G\"{u}rses and G\"{u}rsey \cite{GG} as a specific rotating instance of the algebraically special Kerr-Schild metric:
\begin{equation}\label{GGg}
ds^2=(\eta_{\alpha \beta} +2 H k_\alpha k_\beta) dx^\alpha dx^\beta,
\end{equation}
where $\eta$ is the metric of Minkowski, $H$ is a scalar function and $\vec k$ is a light-like vector both with respect to the spacetime metric and to Minkowski's metric.
Specifically, in Kerr-Schild coordinates $\{\tilde{t},x,y,z\}$ the  G\"{u}rses-G\"{u}rsey metric (\ref{GGg}) corresponds to the choices
\[
H=\frac{\mathcal M (r) r^3}{r^4+a^2 z^2}
\]
and
\[
k_\alpha dx^\alpha =-\frac{r (x dx+y dy) -a (x dy-y dx)}{r^2+a^2}-\frac{z dz}{r}-d\tilde{t},
\]
where $r$ is a function of the Kerr-Schild coordinates implicitly defined by
\begin{equation}\label{defr}
r^4-r^2 (x^2+y^2+z^2-a^2) -a^2 z^2 =0,
\end{equation}
$\mathcal M (r) $ is known as the \textit{mass function}
and the constant $a$ is a rotation parameter.

This metric can be written in Boyer-Lindquist-like coordinates by using the coordinate change defined by
\begin{eqnarray}\label{coordchange}
x+i y&=&(r+i a) \sin\theta\exp\left[i\int(d\phi+\frac{a}{\Delta}dr)\right]\\
z&=&r \cos\theta\\
\tilde{t}&=&t+\int \frac{r^2+a^2}{\Delta} dr-r.
\end{eqnarray}
where now $\Delta=r^2-2 \mathcal M(r) r+a^2$.
The resulting metric takes the form
\begin{equation}\label{gIKerr}
ds^2=-\frac{\Delta}{\Sigma} (dt-a \sin^2\theta d\phi)^2+
\frac{\Sigma}{\Delta} dr^2+\Sigma d\theta^2+\frac{\sin^2\theta}{\Sigma}(a dt-(r^2+a^2)d\phi)^2,
\end{equation}
where $\Sigma=r^2+a^2 \cos^2\theta$.
This metric reduces to the well-known Kerr's solution in B-L coordinates if $\mathcal M(r)=M$=constant. In this case, there is a curvature singularity at $(r=0, \ \theta=\pi/2)$, as can be shown by the divergence of its Kretschmann scalar.

Interestingly, for $a \neq 0 $ and $ \theta \neq \pi/2 $, a surface defined by $ t = $ constant and $ r = 0 $ is free from singularities and has a metric
\[
ds_2^2 = a^2 \cos^2 \theta \, d\theta^2 + a^2 \sin^2 \theta \, d\phi^2 = dx^2 + dy^2,
\]
where the coordinate change $ x \equiv a \sin\theta \cos \phi $, $ y \equiv a \sin\theta \sin\phi $ explicitly shows that the surface is flat. This flat surface corresponds to $ x^2 + y^2 < a^2 $, conventionally referred to as the \textit{disk}. The boundary $ x^2 + y^2 = a^2 $ is termed the \textit{ring}.

Note also the symmetry $\{a,\phi\}\leftrightarrow \{-a,-\phi\}$ in this metric. This allows us, for the sake of simplicity, to assume $a>0$ in this article, since the negative case is covered by using the trivial coordinate change $\phi \rightarrow -\phi$.

In order to analyze the general properties of the superspinar
we will use the following null tetrad-frame:
\begin{eqnarray*}
\mathbf{l} &=&\frac{1}{\Delta} \left( (r^2+a^2) \frac{\partial}{\partial t}+\Delta \frac{\partial}{\partial r}+a \frac{\partial}{\partial \phi}\right),\\
\mathbf n &=&\frac{1}{2 \Sigma} \left( (r^2+a^2) \frac{\partial}{\partial t}-\Delta \frac{\partial}{\partial r}+a \frac{\partial}{\partial \phi}\right),\\
\mathbf m &=& \frac{1}{ \sqrt{2} \varrho } \left(i a \sin\theta \frac{\partial}{\partial t}+\frac{\partial}{\partial \theta}+i \csc\theta \frac{\partial}{\partial \phi} \right),\\
\mathbf{\bar m} &=& \frac{1}{ \sqrt{2} \bar\varrho } \left(-i a \sin\theta \frac{\partial}{\partial t}+\frac{\partial}{\partial \theta}-i \csc\theta \frac{\partial}{\partial \phi} \right),
\end{eqnarray*}
where $\varrho\equiv r+i a \cos\theta$, $\bar\varrho\equiv r-i a \cos\theta$ and the tetrad is normalized as follows $\mathbf l^2=\mathbf n^2=\mathbf m^2=\mathbf{\bar m}^2=0$ and $\mathbf l\cdot \mathbf n=-1= -\mathbf{m}\cdot \mathbf{\bar m}$.
The metric (\ref{gIKerr}) with $\mathcal M(r)\neq 0$ is Petrov type D and the two double principal null directions are $\mathbf l$ and $\mathbf n$ \cite{TorresReg}.

We can also define a real orthonormal basis $\{\mathbf{t}, \mathbf{x}, \mathbf{y}, \mathbf{z } \}$
formed by a timelike vector $\mathbf{t}\equiv (\mathbf{l}+\mathbf{n})/\sqrt{2}$ and three spacelike vectors: $\mathbf{z}\equiv (\mathbf{l}-\mathbf{n})/\sqrt{2}$,  $\mathbf x=(\mathbf m +\bar{\mathbf m})/\sqrt{2}$ and $\mathbf y=(\mathbf m -\bar{\mathbf m}) i/\sqrt{2}$. Then,
$\mathbf t$ and $\mathbf z$ are two eigenvectors of the Ricci tensor with eigenvalue \cite{TorresReg}
\begin{equation}\label{lambda1}
\lambda_1=\frac{2 a^2 \cos^2{\theta} \mathcal M'+r \Sigma \mathcal M''}{\Sigma^2}.
\end{equation}
$\mathbf x$ and $\mathbf y$ are two eigenvectors of the Ricci tensor with eigenvalue
\begin{equation}\label{lambda2}
\lambda_2=\frac{2 r^2 \mathcal M'}{\Sigma^2}.
\end{equation}

The metric has a coordinate singularity in those cases where there are values of $r$ such that $\Delta=0$, which can be removed by a coordinate change (see \cite{TorresCap}).
At $\Delta=0$ the hypersurface $r=$constant becomes light-like and no observer can remain at the specific value for $r$, thus the hypersurface is called a \textit{null horizon}.

The simplest example is Kerr's solution: if $M^2>a^2$ there are two roots: $r^{Kerr}_\pm=M\pm \sqrt{M^2-a^2}$, where the null horizon $r^{Kerr}_+$ is an event horizon, while the null horizon $r^{Kerr}_-$ is a Cauchy horizon. The limiting case $M^2=a^2$ has a degenerate null horizon and the spacetime is called the \textit{extreme} Kerr black hole. If  $M^2<a^2$ there are no roots for $\Delta=0$ and, therefore, no horizons. This is the so-called \textit{hyperextreme} case that provides the model of a classical superspinar.

\subsection{Scalar-Regularity of Superspinars}\label{secRegu}

For a finite $\mathcal{M}(r)$, scalar curvature singularities may exist only where $\Sigma=0$ or, in other words, in $(r=0,\theta=\pi/2)$ (as previously noted for Kerr's solution). Nevertheless, while $\Sigma=0$ is necessary, it is not a sufficient condition to guarantee the existence of curvature singularities.
Conversely, a necessary and sufficient condition for the \emph{absence} of scalar curvature singularities is provided by the following

\begin{theorem}\label{teorema}\cite{TorresReg}
Assuming a metric of type (\ref{gIKerr}) possessing a $C^3$ function $\mathcal M(r)$, all its second order curvature invariants will be finite at $(r=0,\theta=\pi/2)$ if, and only if,
\begin{equation}\label{condisreg}
 \mathcal M (0)= \mathcal M' (0)= \mathcal M'' (0)=0 .
\end{equation}
\end{theorem}

Note that any function admitting an expansion around $r=0$ in the form $\mathcal M(r)=c r^n + O(r^{n+1})$, with $c$ a constant and $n\geq 3$, will satisfy the conditions. (This will be written here also as $\mathcal M(r)= O( r^n )$ or $\mathcal M(r)\sim r^n$). Consider the examples of the mass functions provided by Bonanno and Reuter (BR) \cite{B&R}\cite{R&T}\cite{TorresExt} and by Hayward \cite{Hay2006} (that will be useful later to illustrate different properties of regular superspinars)
\begin{equation}\label{B&H}
\mathcal{M}(r)_{BR}=\frac{m r^3}{r^3+l^2 (r+L)}
 \ \ ; \ \ \mathcal M(r)_{Hayward}=\frac{m r^3}{r^3+l^3},
\end{equation}
where $l$ and $L$ are constants. These mass functions admit the following expansions around $r=0$:
\[
\mathcal M(r)_{BR}=\frac{m}{l^2 L} r^3 + O(r^4)
 \ \ ; \ \ \mathcal M(r)_{Hayward}=\frac{m}{l^3} r^3 + O(r^6).
\]
In the specific BR case, computations in the Quantum Einstein Gravity framework lead the authors to specific values for the constants with $l\sim l_{Planck}$ and $L\sim$ Schwarzschild`s radius. In this way, the deviations from the classical case will be relevant for $r$ of the order of some Planck lengths. This is not specific to this example, but a common result and a common assumption in the case of heuristic models (such as Hayward's model, $l\sim l_{Planck}$).

The reader can consult other mass functions in the literature with the following approaches:
In the framework of Conformal Gravity  \cite{BMR},
in the framework of Shape Dynamics in \cite{G&H},
inspired by Supergravity in \cite{Buri},
by Loop Quantum Gravity in \cite{C&M}
and by non-commutative gravity in \cite{S&S}.

Note that the above theorem allows to control the specific case of \emph{scalar curvature singularities}, which arguably are the most serious type of curvature singularities. However, since scalar polynomials do not fully characterize the Riemann tensor, it does not cover the possibility of the existence of curvature singularities with respect to a parallelly propagated basis (\textit{p.p. curvature singularity})\cite{H&E}. To date, this possibility has not been thoroughly explored in the literature. Additionally, conical singularities may also exist (see next subsection). Taking these caveats into account, we are \emph{not} following the common practice in the field of referring to superspinars satisfying the given theorem as \textit{regular}, but instead we are using the term \textit{scalar-regular} (SR).

\subsection{Regarding the Potential Non-Existence of Extensions Through the Disk}\label{secr0}

The curves that reach the disk of a superspinar are reaching a regular point. In Kerr's case, and due to a lack of differentiability at the disk, an analytic extension of the spacetime has to be made through $r=0$ in order to continue the curves. This is accomplished by allowing the coordinate $r$ to take negative values \cite{H&E}. However, this analytic continuation also introduces some of the issues associated with Kerr's solutions: The $r<0$ extended spacetime can be seen as a negative mass spacetime and causality violations manifest in the extended spacetime \cite{Carter1968a}.
In contrast, the following result has been demonstrated for the SR case \cite{TorresInt} (summarized in the appendix):

\begin{em}
SR rotating models with metric (\ref{gIKerr}) (without an extension through $r=0$) have a high degree of differentiability at the disk (at least $C^n$, with $n \geq 3$).
\end{em}

In this way, SR rotating objects do not have differentiability problems at the disk and an extension through $r=0$ with $r<0$ is \emph{not needed}. In other words, a spacetime with metric (\ref{gIKerr}) and mass function satisfying the conditions for scalar regularity can be topologically trivial without the need for a physical reinterpretation of the disk. (See, by contrast, the case of the Kerr metric in \cite{IsraelThin}\cite{Israel}\cite{Hamity}).

Note that the key here is that a metric alone does not determine the global topology of the spacetime.
It would be a different matter if extra information was added. For example, if in the spirit of \cite{G&V}, we consider SR-superspinars that were the limiting case of non-topologically trivial Kerr superspinars that evolve their mass function until reaching the conditions for absence of scalar curvature singularities. In such a case, one should logically consider a non-trivial topology for SR-superspinars, i.e., negative values for the coordinate $r$. This point of view is mathematically consistent. It only requires the presence of conical singularities at the ring \cite{Maeda} and a non-straightforward continuation of the mass function for negative values of $r$ \cite{TorresExt}. As commented, the physical problems with this approach are that the $r<0$ extended spacetime can be seen as a negative mass spacetime and causality violations manifest in the extended spacetime

In this article, we will mainly take the approach of considering that the global spacetime of SR-superspinars is topologically trivial, but we will also comment on the observational consequences of extending the metric for negative values of $r$.

%As a consequence, SR superspinars with non-negative mass functions will not be able to act repulsively.

\subsection{Causality}\label{secCaus}

In general, it is reasonable to require a time-orientable spacetime to be free of closed causal curves, as their existence would lead to logical paradoxes. A spacetime devoid of closed causal curves is referred to as \textit{causal} \cite{H&E}. Furthermore, if no closed causal curves appear even under any small perturbation of the metric, the spacetime is termed \textit{stably causal}. As mentioned, it is well-known that the standard analytical extension of the Kerr metric is non-causal.
However, for SR rotating objects it can be shown the following
\begin{theorem} \cite{Maeda}\cite{TorresCap}
If $r \mathcal M(r)\geq 0$ for all $r$, then the rotating model with metric (\ref{GGg}) [or (\ref{gIKerr})] will be stably causal.
\end{theorem}
Note that, as a consequence, for an SR superspinar (unextended through $r=0$), a non-negative mass function suffices to guarantee a stably causal spacetime.

\subsection{No horizon condition and global structure of the spacetime}\label{Horizons}

\begin{figure}[htp]
\includegraphics[scale=.7]{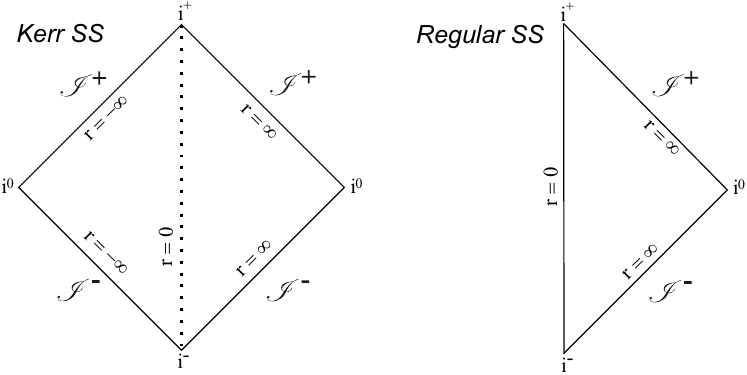}
\caption{\label{PenroseK&RSS} \emph{Left}: Penrose diagram for Kerr's superspinar. The spacetime is extended through $r=0$ to an asymptotically flat region with negative values for $r$. (Note that the diagram is valid for $\theta\neq \pi/2$. The diagram with $\theta= \pi/2$ will require drawing the ring singularity). \emph{Right}: Penrose diagram for a topologically trivial SR superspinar. (Note that, since there are no singularities, the diagram is valid for all $\theta$).}
\end{figure}

A superspinar is characterized by the absence of horizons (for all $r$), which necessitates the condition
\begin{equation}\label{eqdelta}
\Delta=r^2-2 \mathcal M(r) r+a^2>0,
\end{equation}
implying that $r$ is a spacelike coordinate throughout the entire spacetime. This simplifies obtaining the global structure of the spacetime starting from BL coordinates.
It is important to note that the causal structure of a Kerr superspinar includes an extended region with negative values for $ r $. In contrast, for an SR superspinar, there is no need for an extension through $ r = 0 $. The Penrose diagrams for both cases are shown in Figure \ref{PenroseK&RSS}.

As suggested by the figure, the absence of an event horizon in superspinars is significant, as it seems to imply that a distant observer could receive information from the inner high-curvature regions near $ r = 0 $. In principle, this could be used to observationally test various approaches to Quantum Gravity. This possibility will be explored in Section \ref{secPheno}.

\section{Effective energy-momentum and Planckian region}\label{ecs}

Even if here we are not confined to General Relativity, we can
consider the existence of an \textit{effective energy-momentum tensor} defined through
\[
T_{\mu\nu}\equiv R_{\mu\nu}-\frac{1}{2} \mathcal R g_{\mu\nu}.
\]

One can write $\mathbf{T}$ explicitly for a superspinar as \cite{GG}\cite{TorresReg}
\[
T_{\mu\nu}=-\lambda_2 (-t_\mu t_\nu + z_\mu z_\nu)- \lambda_1 (x_\mu x_\nu+ y_\mu y_\nu).
\]

The (effective) density being $\mu=\lambda_2$ and the (effective) pressures being $p_x=p_y=-\lambda_1$ and $p_z=-\lambda_2$ (see (\ref{lambda1}) and (\ref{lambda2})).

\begin{figure}[ht]
\includegraphics[scale=.8]{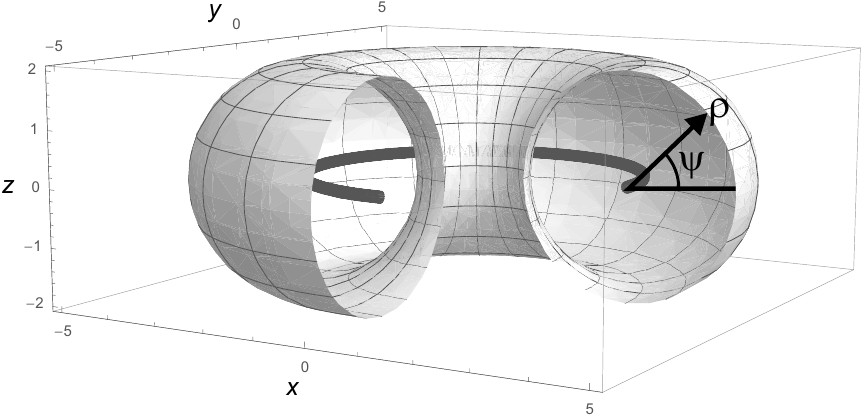}
\caption{\label{torusring} A plot of the behaviour of the toroidal coordinates with the (SR) ring ($\rho=0$) highlighted in black.}
\end{figure}

It seems natural that any deviations of a realistic SR superspinar from its classical counterpart should be concentrated in a region around the ring. To rigorously analyze this intuition, let us study the ring in greater depth by introducing a set of coordinates better adapted to its structure. These coordinates will help us examine the directional behavior of physical quantities around the ring: the \textit{toroidal} coordinates $\{\rho,\varphi,\psi\}$ (see figure \ref{torusring}). These coordinates are related to the Kerr-Schild coordinates through
\begin{eqnarray}\label{transf}
  x &=& (a+\rho \cos \psi) \cos \varphi,\nonumber \\
  y &=& (a+\rho \cos \psi) \sin \varphi, \\
  z &=& \rho \sin \psi. \nonumber
\end{eqnarray}
Note that in these coordinates the ring is defined by $\rho=0$.

From the relationship between K-S and B-L coordinates, it follows that the coordinates $\{\rho,\psi\}$ are related to the Boyer-Lindquist coordinates $\{r,\theta\}$ through
\begin{eqnarray}\label{rcosth}
  r \cos \theta &=& \rho \sin \psi, \\
  (r^2+a^2) \sin^2\theta &=& (a+\rho \cos \psi)^2.\nonumber
\end{eqnarray}

This implies that, for a topologically trivial SR-superspinar, around the ring

\begin{eqnarray}\label{rthaprox}
  r &\simeq& \sqrt{a (1+\cos \psi) \rho},\\
  \cos \theta &\simeq& \frac{\rho^{1/2} \sin \psi}{\sqrt{a (1+\cos\psi)}}.\nonumber
\end{eqnarray}

Let us now analyze the dependence of the effective density and pressures on $\rho$ in a region around the ring for a mass function $\mathcal M(r)\sim r^n,\ (n\geq 3)$ around $r=0$. With the help of (\ref{rthaprox}):

\begin{eqnarray*}
\mu=-p_z&=&\lambda_2=\frac{2 r^2 \mathcal M'}{\Sigma^2} \sim \rho^\frac{n-3}{2}\\
p_x=p_y&=&-\lambda_1=-\frac{2 a^2 \cos^2{\theta} \mathcal M'+r \Sigma \mathcal M''}{\Sigma^2} \sim \rho^\frac{n-3}{2}.
\end{eqnarray*}

In this way, the effective density and pressures vanish at the ring for $n > 3$. On the other hand, for $n=3$ the densities and pressures are finite functions of the direction of approach to the ring $\psi$. Only in this case the density and pressures are not continuous at the ring.

Specifically, for $\mathcal M(r)\equiv m_3 r^3+O(r^4)$ with $m_3>0$ a constant,

\begin{eqnarray}\label{mupn3}
\mu=-p_z&=&\lambda_2= 6 m_3 \cos^4 (\psi/2),\\
p_x=p_y&=&-\lambda_1=3 m_3 \cos^2 (\psi/2) (\cos \psi-3).\nonumber
\end{eqnarray}

\begin{figure}[ht]
\includegraphics[scale=.7]{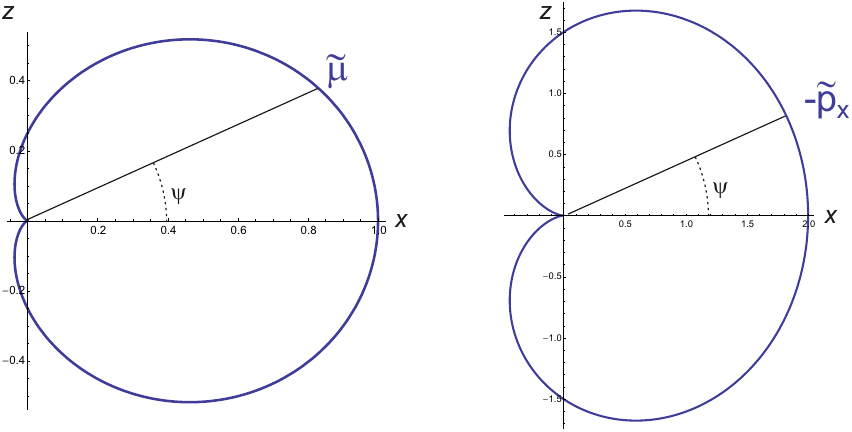}
\caption{\label{denspres} Polar plot of $\tilde\mu\equiv \mu/(6 m_3)$ and (minus) $\tilde p_x\equiv p_x/(3 m_3)$ around the ring for the case $\mathcal M(r)\sim r^3$ around $r=0$.}
\end{figure}

For both the effective density and pressures, the value zero is reached in the direction of the disk $\psi=\pi/2$ and the maximum absolute value is reached in the opposite direction  $\psi=0$.
Polar plots of the effective density and pressures for this particular case are shown in figure \ref{denspres}.

Note that the disk (and the ring when approached with $\psi=\pi$) can be described as \textit{minkowskian} ($\mu=p_x=p_y=p_z=0$).

\begin{figure}[ht]
\includegraphics[scale=.9]{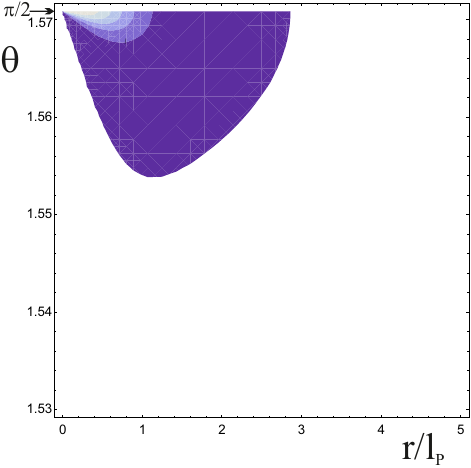}
\caption{\label{QGdensity} An example of the region with planckian densities ($\mu\sim \mu_{Planck}$) in BL coordinates (dark region). Note that the region has a maximum minor radius of the order of a few $l_P$ and that it is concentrated around $\theta=\pi/2$ (i.e., the equatorial plane). In this example it has not been necessary to plot the region with $0\leq\theta\leq 1.53$ since it does not have planckian densities. We have specifically used the BR model. However, the results are qualitatively similar for all SR models under the assumptions in subsection \ref{secRegu}.}
\end{figure}

We call \textit{planckian region} the region in which the curvature, effective density and effective pressures reach their corresponding high Planckian values. In the literature, this region is  sometimes assumed to be an ellipsoid containing the disk \cite{B&F}. Here, we will locate it using the assumptions in subsection \ref{secRegu}. To accomplish this, it suffices to compute the values of the effective density and pressures from the values of $\lambda_1$ and $\lambda_2$ in BL coordinates. (See figure \ref{QGdensity}).
From this, we can deduce that the planckian region is not ellipsoidal. Moreover, it is not even isotropic around the ring, when projected in a plane perpendicular to the x-y plane and containing the z-axis. In other words, the planckian region does not have the shape of a torus, but rather what we will call a \textit{pseudotorus}. The planckian pseudotorus has a \textit{major radius} corresponding to the ring of size $|a|$, and it has a \textit{mean minor radius} of the order of a few $l_P$ [with the smallest radius (0) in the direction of the disk ($\psi=\pi$) and the largest radius in the direction of the equatorial plane ($\psi=0$)]. As a consequence, the ratio between the major and the mean minor radius will satisfy
\[
\frac{|a|}{l_p}\gtrsim \frac{M}{M_P},
\]
where $M_P$ is Planck's mass.
Thus, the major radius of the pseudotorus will be immensely large compared to the mean minor radius, except when dealing with superspinars of Planckian mass. For example, a superspinar with a mass twice that of our Sun will have a major radius greater than 3 km, which is approximately $10^{38}$ times larger than the mean minor radius of the Planckian region. This implies that, under these assumptions, an \textit{observer} could pass through the (Minkowskian) disk of an astrophysical superspinar without ever traversing a true planckian region.

\section{Null geodesics reaching $r=0$}\label{ngd}

The trajectory of the photons
in the spacetime of an SR superspinar is described by an action $S=S(x^\alpha)$. The momentum of the photons is
\[
p_{\mu}\equiv \frac{\partial S}{\partial x^\mu}
\]
and satisfies
\begin{equation}\label{photon}
g^{\alpha\beta} p_\alpha p_\beta=0.
\end{equation}
The stationary and axisymmetric nature of the spacetime described by the metric (\ref{gIKerr}) implies two conserved quantities in the trajectory of the photon: the energy $E\equiv -p_t$ and the angular momentum $L\equiv p_\phi$. If there is a separable solution for $S$, by using the definition of the momentum, we could rewrite it as
\[
S=-E t+L \phi+ S_r+ S_\theta,
\]
where we have introduced the new functions $S_r=S_r(r)$ and $S_\theta=S_\theta(\theta)$. In this way, (\ref{photon}) can now be written as
\begin{equation}\label{SrSth}
-\Delta \left(\frac{d S_r}{dr} \right)^2+ \frac{[(r^2+a^2) E-a L]^2}{\Delta}=\left(\frac{d S_\theta}{d\theta} \right)^2+ \frac{(L-a E \sin^2\theta)^2}{\sin^2\theta}.
\end{equation}
In this equation, the left-hand side depends only on $r$, while the right-hand side depends only on $\theta$. In this way, they define a constant which we will denote by
\begin{equation}\label{KSth}
K=\left(\frac{d S_\theta}{d\theta} \right)^2+ \frac{(L-a E \sin^2\theta)^2}{\sin^2\theta}.
\end{equation}
From $dx^\mu/d\lambda=p^\mu=g^{\mu\nu} p_\nu$ and using (\ref{SrSth}) and (\ref{KSth}) one gets (see, for instance, \cite{Tsuka})
\begin{eqnarray}
\Sigma \frac{dt}{d\lambda}&=&-a (a E \sin^2\theta-L) +\frac{(r^2+a^2) P(r)}{\Delta},\label{dtdl}\\
\Sigma \frac{dr}{d\lambda}&=&\sigma_r \sqrt{R(r)},\label{drdl}\\
\Sigma \frac{d\theta}{d\lambda}&=&\sigma_\theta \sqrt{\Theta(\theta)},\label{dthdl}\\
\Sigma \frac{d\phi}{d\lambda}&=&-\left( a E-\frac{L}{\sin^2\theta}\right) +\frac{a P(r)}{\Delta}\label{dphdl}
\end{eqnarray}
where
\begin{eqnarray}
P(r)&\equiv& E (r^2+a^2)-a L,\\
R(r)&\equiv& P(r)^2-\Delta [(L-a E)^2+\mathcal Q],\\
\Theta(\theta)&\equiv&\mathcal Q+\cos^2\theta \left(a^2 E^2-\frac{L^2}{\sin^2\theta}\right).
\end{eqnarray}
with $\mathcal Q\equiv K-(L-a E)^2$ the Carter constant and $\sigma_r$ and $\sigma_\theta$ can take the values $\pm 1$ independently.

$R$ can also be rewritten as
\begin{equation}\label{deffR}
R/E^2=r^4+(a^2-\xi^2-\eta) r^2 + 2 \mathcal M (r) [(\xi-a)^2+\eta] r-a^2 \eta,
\end{equation}
where $\eta\equiv \mathcal Q/E^2$ and $\xi\equiv L/E$.

For the null geodesics traversing the disk, $r=0$ must be reached for some value of their affine parameter. For a finite mass function, $R$ will take the value $R(r=0)=-a^2 E^2 \eta$. According to (\ref{drdl}), $R\geq 0$, what implies as condition for the null geodesics reaching the disk or the ring: $\eta\leq 0$.

\subsubsection{The case $\eta <0$}

Due to the symmetry, we will consider here that null geodesics approaching $r=0$ from $0\leq \theta \leq \pi/2$.

The function $\Theta$ can be written as
\begin{equation}\label{ThetaEq}
\frac{\Theta}{E^2}=\eta+(a-\xi)^2-\left(a \sin \theta-\frac{\xi}{\sin \theta}\right)^2
\end{equation}
which implies that a necessary condition for the existence of the solutions is
\begin{equation}\label{ineqC}
|\eta|\leq (a-\xi)^2.
\end{equation}

$\Theta(\theta=\pi/2)=E^2 \eta<0$ and this is not possible according to (\ref{dthdl}). Thus, there will always be a maximum value for $\theta=\theta_{max}$ which cannot be traversed. As a consequence, these geodesics can not reach the ring ($\theta=\pi/2$).

Two possibilities can be considered according to the value of $\xi$:
\begin{itemize}
\item If $\xi \neq 0$, $\Theta$ is necessarily negative around $\theta=0$. Therefore the null geodesics with
    $\xi \neq 0$ will not cross the disk near the z-axis as there will also be a minimum value for $\theta=\theta_{min}>0$. See figure \ref{disk}. The limits allowed for $\theta$ will be the solutions of $\Theta=0$ that, writing $\mu\equiv \cos\theta$, are simply the solutions of
\[
\eta-(\xi^2+\eta-a^2) \mu^2-a^2 \mu^4=0.
\]
The existence of the solutions (non-negative discriminant and $0\leq \mu^2<1$) imposes that the geodesics crossing the disk must also satisfy $|\eta|\leq a^2$ and $|a|-\sqrt{|\eta|}\geq |\xi|$.

\item If $\xi=0$, according to (\ref{ineqC}) the geodesic should satisfy $|\eta|\leq a^2$. The requirement that $\Theta\geq 0$ imposes a limit (as discussed above) that in this case takes the explicit form $\cos^2\theta_{max}= |\eta|/a^2$. However, it does not impose a minimum value for $\theta$, implying that in this case the geodesics can cross the disk near or at the $z$ axis.

\begin{figure}[h]
\includegraphics[scale=.7]{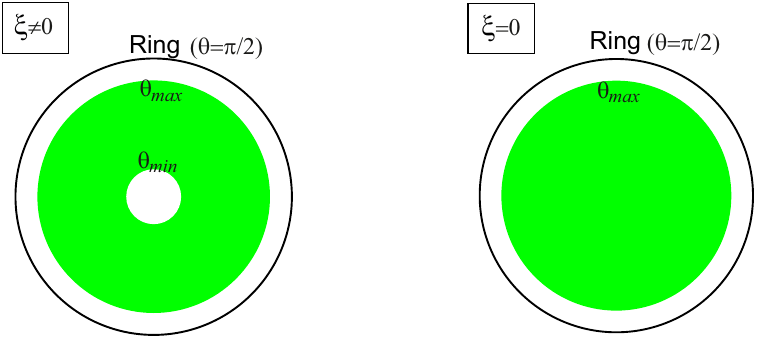}
\caption{\label{disk} The disk of a superspinar bounded by the ring (in the plane $x-y$ of K-S coordinates) with the allowed regions (green) for the null geodesics crossing it in the case $\xi\neq 0$ and $\xi=0$.}
\end{figure}

\end{itemize}

\begin{figure}[h]
\includegraphics[scale=1]{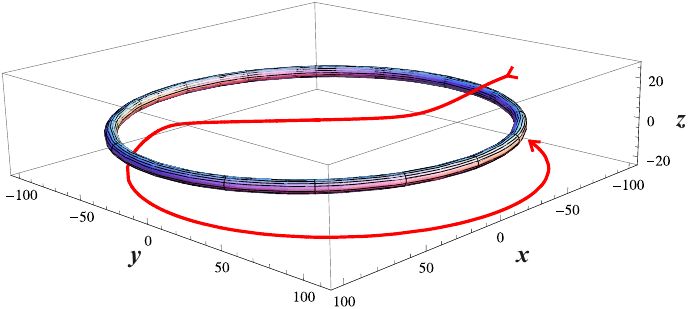}
\caption{\label{xino0} A light-like geodesic (red) with non-null angular momentum ($\xi\neq 0$) crossing the disk of an SR-superspinar from the region with $z>0$ to the region with $z<0$. The planckian region is represented by a pseudotorus. As required in order to cross the disk, Carter's constant is negative $\eta<0$ and the parameters have been chosen among the required conditions for these geodesics. Specifically, $\eta=-10$, $\xi=1$ and $E=1$. On the other hand, the BR superspinar is defined by the parameters $M=100$, $a=100$, $l=1$ and $L=100$ that, as required for superspinars, avoid the formation of horizons since $\Delta\neq 0$. In this topologically trivial SR-superspinar, once the geodesic crosses the disk, $r$ monotonically increases along it.}
\end{figure}

\begin{figure}[h]
\includegraphics[scale=1]{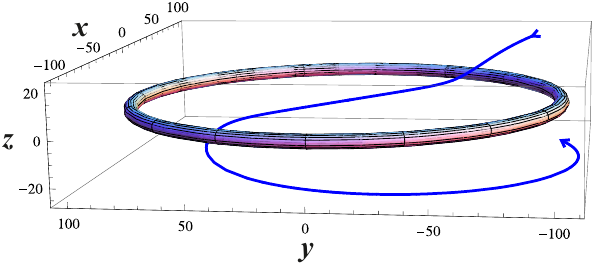}
\caption{\label{xi0}  A light-like geodesic (blue) with null angular momentum ($\xi=0 $) crossing the disk of an SR-superspinar from the region with $z>0$ to the region with $z<0$. The parameters have been chosen among the required conditions for these geodesics. Specifically, $\eta=-10$, $\xi=0$ and $E=1$. On the other hand, the superspinar is the same as in figure \ref{xino0}. As can be seen, the main difference with the previous figure is that the region around the center of the disk is accessible by the null geodesic with $\xi=0$. In fact, in the specific case shown, it crosses the disk at $(r=0,\theta=0)$. }
\end{figure}

The explicit 3-dimensional image of a geodesic crossing the disk of a topologically trivial SR superspinar in Kerr-Schild coordinates with a parameter $\xi\neq 0$ is illustrated in figure \ref{xino0}, while the case of a geodesic crossing the disk with a parameter $\xi=0$ is illustrated in figure \ref{xi0}. In order to obtain these figures, the system of differential equations (\ref{drdl})+(\ref{dthdl})+(\ref{dphdl}) has been solved with some initial conditions, then the coordinate change (\ref{coordchange}) has been used to get $\{x(\lambda),y(\lambda),z(\lambda)\}$. These figures emphasize the fact that, contrary to Kerr's solutions, in the topologically trivial SR case the null geodesics do not cross to a new spacetime with negative $r$. Along their trajectory, first $r$ decreases until reaching the disk ($r=0$), and then $r$ increases as the null geodesic moves away from the disk.

\subsubsection{The case $\eta=0$} \label{eta0}

For null geodesics not restricted to the equatorial plane, the fact that $\Theta\geq 0$ imposes
\[
\xi^2\leq a^2
\]
and force them to be in the range $\theta_{min}\leq\theta\leq \pi-\theta_{min}$, where $\theta_{min}=\arcsin|\xi/a|$.

As noted previously, only null geodesics with $\eta=0$ have the potential to reach the ring. To examine this possibility, observe that $\Theta(\theta=\pi/2)=0$.
The system (\ref{drdl})+(\ref{dthdl}) can be rewritten by defining a new parameter $s$ such that $d\lambda=\Sigma(s) ds$ as
\begin{eqnarray*}
\frac{dr}{d s}&=&\sigma_r \sqrt{R(r(s))},\label{drds}\\
\frac{d\theta}{d s}&=&\sigma_\theta \sqrt{\Theta(\theta(s))}.\label{dthds}\\
\end{eqnarray*}

Therefore, $(r=0,\theta=\pi/2)$ is an isolated critical point of this system. The behaviour of the critical point is very different in Kerr's superspinars or SR-superspinars:
\begin{itemize}
\item In Kerr's superspinars as the isolated critical point is approached
    $\partial \sqrt{R}/\partial r \sim r^{-1/2}$.
    Thus, this diverges at the ring and the system can not be linearized.
     Practically all null geodesics approaching $r=0$ reach the disk
     ($\theta<\pi/2$) (see figure \ref{kerrring}), where they find a turning point ($dr/d\lambda=dr/ds=0$) and they cannot traverse the disk towards negative $r$, according to the condition $R>0$ and the fact that, around $r=0$,
     $R\simeq 2 M r (a-\xi)^2$. This leads us to the important result that only null geodesics approaching the disk from the equator ($\theta(\lambda)=\pi/2=$constant) can reach or emerge from the ring singularity. As a consequence, the ring singularity is relatively well concealed: only observers in the equatorial plane can observe it.

\begin{figure}[h]
\includegraphics[scale=1]{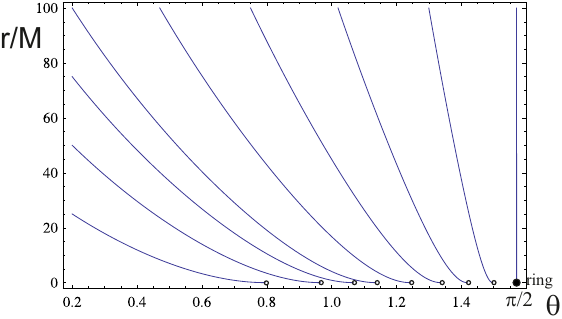}
\caption{\label{kerrring}  A numerical integration of the system (\ref{drdl})+(\ref{dthdl}) for Kerr's case
($\mathcal M(r)=$constant) and null geodesics satisfying $\eta=0$ showing that only null geodesics approaching the disk from the equator ($\theta(\lambda)=\pi/2=$constant) are able to reach the ring. The other geodesics reach the disk ($r=0,\theta<\pi/2$), where they find a turning point. Conversely, null geodesics from the ring of the Kerr superspinar will only reach faraway observers in the equatorial plane. In this particular plot $a/M=2$ and $\xi=0$, but other parameters provide the same \emph{qualitative} figure.}
\end{figure}

\item For SR-superspinars one has, at the isolated critical point, $\partial \sqrt{R}/\partial r =\partial \sqrt{\Theta}/\partial \theta=\sqrt{a^2-\xi^2}\ (\in \Re)$.
    Thus, the system is linearizable. In this way, it is easy to check that all null geodesics coming out of $r=0$ do so from the critical point, i.e., the ring (see figure \ref{regucrit}). Thus, contrary to Kerr's case, there are null geodesics from the ring to distant observers outside the equatorial plane. We will delve into its observational consequences in the next section.

\begin{figure}[h]
\includegraphics[scale=1]{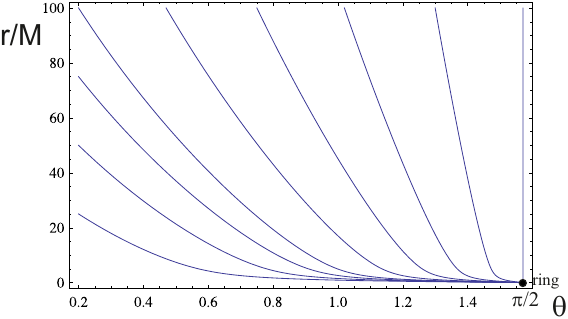}
\caption{\label{regucrit}  A numerical integration of the system (\ref{drdl})+(\ref{dthdl}) for an SR-superspinar showing that all null geodesics with $\eta=0$ approaching $r=0$ go to the ring (and, conversely, null geodesics from the ring will reach faraway observers even if they are outside the equatorial plane). The parameters defining the BR superspinar have been chosen $l/M=10^{-2}, L/M=1$ and $a/M=2$. The null geodesics have $\xi=0$, but other values provide the same \emph{qualitative} figure.}
\end{figure}

\end{itemize}

\section{The silhouette of superspinars}\label{secPheno}

\begin{figure}[ht]
\includegraphics[scale=.7]{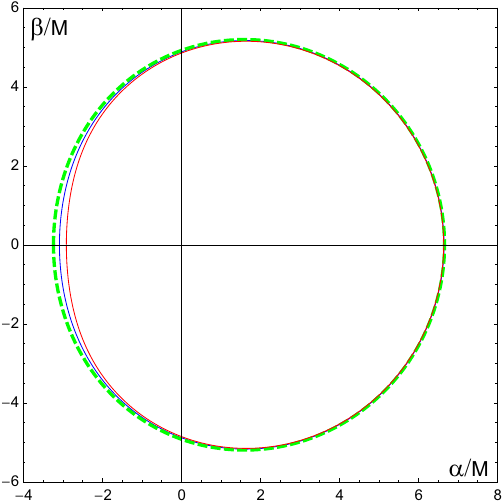}
\caption{\label{shadows} Shadows of rotating black holes (RBH) are shown for three different cases. We have fixed the rotation parameter to $a/M = 0.8$ and the inclination angle to $\theta_i = \pi/2$ (i.e., it is viewed from the equatorial plane). The dashed green boundary of the shadow corresponds to the Kerr black hole case. The blue boundary of the shadow corresponds to Hayward's RBH with $l/M = 0.5$ \cite{TorresInt}. The red boundary of the shadow corresponds to Hayward's RBH with $l/M = 0.6$. Note that for smaller values of $l/M$, the deviations from Kerr's shadow are smaller. As expected, when $l$ is on the order of the Planck mass, the deviation of the shadow from the classical case is negligible.}
\end{figure}

Despite the similarities between black holes and superspinars, their images, as seen by faraway observers, are very different.
For black holes, a defining characteristic is the existence of an event horizon. To a distant observer, photons coming from the background of a black hole and passing through the unstable circular orbits around it will eventually cross the event horizon and, therefore, will not be able to reach the distant observer. This phenomenon casts a relatively large \emph{shadow} for distant observers.
There are numerous studies on shadows of rotating black holes (RBH) in different frameworks. See, for example,
\cite{AAAG,ASG,Am&G,E&H,LGPV,L&B,S&S0,BCY,HGE}.
In order to illustrate the typical results obtained, Figure \ref{shadows} shows the shadows of two SR-RBHs compared to the shadow of a Kerr RBH\footnote{The computations we have carried out to plot this figure are the same as the ones we are going to use to obtain the silhouette of superspinars.}.

Superspinars do not have an event horizon and, therefore, there are no photons crossing it and causing a shadow. Nevertheless, there are related phenomena associated with superspinars due to the existence of unstable circular orbits and the potential visibility of their ring, which will produce a characteristic \emph{silhouette}.

Let us start with the unstable circular orbits. Some null geodesics coming from the background and passing near an unstable circular orbit will be unable to reach a distant observer on the other side. Since the retrograde
photon orbits are less affected by the frame dragging of the superspinar, this phenomenon can cause an open \emph{arc} structure that opens to the left for $a>0$ (or to the right for $a<0$).
Equation (\ref{drdl}) implies that there would be  unstable circular orbits at a certain $r=r_0$ whenever $R(r_0)=R'(r_0)=0$ and $R''(r_0)>0$.
To exploit this, one has to follow the expected steps: Consider the equation $R(r_0)=0$ from eq.($\ref{deffR}$) and
take the derivative of $R$ with respect to $r$, which yields
\[
R'/E^2=4 r^3+2 (a^2-\xi^2-\eta) r+ 2 \mathcal M (r) [(\xi-a)^2+\eta] f(r),
\]
where
\[
f(r)\equiv 1+\frac{r \mathcal M' }{\mathcal M}.
\]
Using the conditions for an unstable orbit one gets the quadratic equation with respect to $\xi$:
\begin{align}\label{quadratic}
a^2 &(r_0-f_0 \mathcal M_0)\xi^2-2 a \mathcal M_0 [(2-f_0) r_0^2-f_0 a^2] \xi-r_0^5+\\
 &+(4-f_0) \mathcal M_0 r_0^4-2 a^2 r_0^3+2 a^2 \mathcal M_0 (2-f_0) r_0^2 -a^4 r_0-a^4 \mathcal M_0 f_0=0,\nonumber
\end{align}
where $\mathcal M_0\equiv\mathcal M(r_0)$ and $f_0\equiv f(r_0)$.

The solution of the quadratic equation (\ref{quadratic}) with the plus sign simply provides
\[
\xi_+=\frac{a^2+r_0^2}{a},
\]
that implies
\[
\eta_+=-\frac{r_0^4}{a^2},
\]
and $R''(r_0)=8 E^2 r_0^2>0$.
On the other hand, these orbits can only exist if, according to (\ref{dthdl}), $\Theta\geq 0$. Since in this case we have $\eta_+ +(a-\xi_+)^2=0$, from (\ref{ThetaEq}), it follows that $\Theta \propto -(a \sin{\theta} -\xi_+/\sin \theta)^2$. Thus, the existence of such orbits requires a constant $\theta$ that satisfies $a \sin^2 \theta=\xi_+$, or equivalently, $\sin^2\theta=1+r_0^2/a^2$. This admits a unique solution: $\{r_0=0,\theta=\pi/2\}$, which corresponds to the ring.

The solution of the quadratic equation (\ref{quadratic}) with the minus sign provides us with
\[
\xi_-\equiv\frac{4 \mathcal M_0 r_0^2-(r_0+f_0 \mathcal M_0)(r_0^2+a^2)}{a (r_0-f_0 \mathcal M_0)}.
\]
that implies
\[
\eta=\eta_-\equiv \frac{r_0^3 [4 (2-f_0) a^2 \mathcal M_0-r_0 [r_0-(4-f_0)\mathcal M_0]^2]}{a^2 (r_0-f_0 \mathcal M_0)^2}.
\]
This leads us, depending on the values of $r_0$, both to unstable orbits ($R''(r_0)>0$), that will be visible to faraway observers, and also to stable orbits ($R''(r_0)<0$) that will be treated in section \ref{SCOs}.

Consider now an observer at a large distance from the rotating object. The observer views the object with an inclination $\theta_i$ relative to the $z$-axis of the superspinar. The curve formed on the sky by the unstable orbits can be expressed in celestial coordinates $\alpha$ and $\beta$, using $r_0$ as its parameter, as follows \cite{Tsuka}
\begin{equation}\label{celestial}
\alpha=\frac{-\xi_-}{\sin\theta_i}\hspace{.5 cm}; \hspace{.5 cm} \beta= \pm \sqrt{\eta_-+(a-\xi_-)^2-\left(a \sin \theta_i-\frac{\xi_-}{\sin\theta_i} \right) ^2} .
\end{equation}

\subsection{Observing the unstable circular orbits: Classical vs SR case}

In the case of a classical superspinar there are three defining observational characteristics \cite{Vries2000}\cite{H&Maeda}:
\begin{enumerate}
\item Using the eqs.(\ref{celestial}) one finds that the unstable photon orbits with positive $r_0$ form an (open) \emph{arc} in the sky.

\item Using the eqs.(\ref{celestial}) one finds that the unstable photon orbits with negative $r_0$ form a closed curve with a dark interior, the \emph{dark spot}, which is due to the geodesics crossing the disk and never returning to the region with positive $r$.

\item The geodesics with $\theta$ constant ($\neq \pi/2$) along them (which are principal null geodesics) traverse the disk and never return to the region with positive $r$. In this way, observers with an inclination $\theta_i$ do not receive the light of the geodesic with $\theta=\pi-\theta_i$. This \emph{dark point} is rather irrelevant from an observational point of view, even more if one takes into account that it would be hidden by the dark spot, except if the observer is precisely situated in the equatorial plane of the superspinar.
\end{enumerate}

\begin{figure}[htp]
\includegraphics[scale=.7]{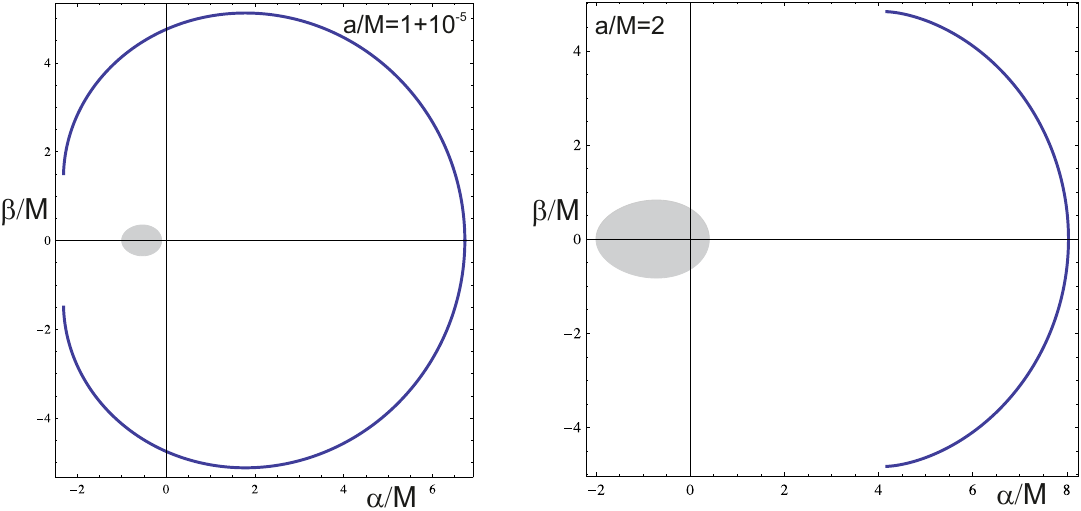}
\caption{\label{KSS} The silhouette of a Kerr superspinar characterized by an arc (blue) (due to the existence of unstable photon orbits of constant $r$ ($>0$)) and a dark spot (grey) (due to the unstable photon orbits with negative $r$). The parameters defining the superspinar have been chosen $a/M=1+10^{-5}$ (left) $a/M=2$ (right). The inclination angle of observation in both cases is $\theta_i=\pi/3$.}
\end{figure}

Examples of silhouettes of Kerr superspinars are shown in Figure \ref{KSS} with the same relationships $a/M$ as in the examples in \cite{H&Maeda}.

\begin{figure}[htp]
\includegraphics[scale=.7]{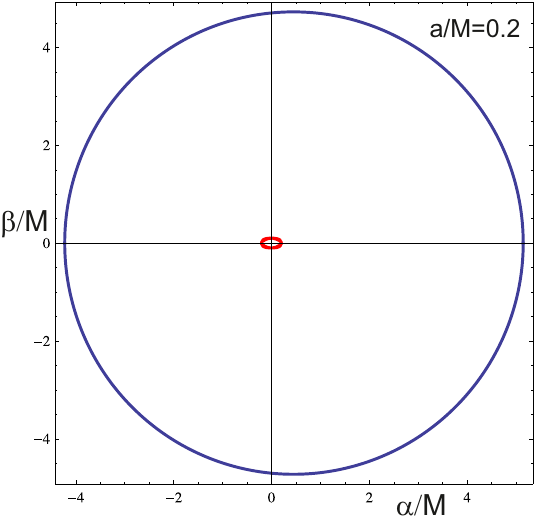}
\caption{\label{B&RSSclosed} The silhouette of a topologically trivial SR-superspinar in which quantum effects reach a radius of order $M$, and $a/M$ is small enough, has a \textit{closed arc} (blue) (due to the existence of unstable photon orbits of constant $r$). Practically the whole region inside the closed curve would be dominated by quantum gravity effects.  In this figure the ellipse highlighted in red is now just the position of the ring. The specific parameters defining the BR superspinar have been chosen $l/M=0.7, L/M=1$ and $a/M=0.2$. The inclination angle of observation is $\theta_i=\pi/3$.}
\end{figure}

In the case of a topologically trivial SR-superspinar there is no region with negative $r$. In this way, there cannot be a dark spot or a dark point. There are two defining observational characteristics:
\begin{enumerate}
\item Using the eqs.(\ref{celestial}) one finds that the unstable photon orbits with constant $r$ form an \emph{arc}. Under the assumptions in subsection \ref{secRegu} this arc will be open, as in the classical case. (However, If we skip those assumptions and let the planckian region to have a size $l\sim M$  this arc could also be closed. See figure \ref{B&RSSclosed}).
\item There are no null geodesics disappearing at a ring singularity, instead there is an SR ring. As a consequence, the planckian region around the ring could emit information that would reach distant observers. This possibility is specifically treated in the next subsection.
\end{enumerate}

\begin{figure}[htp]
\includegraphics[scale=.7]{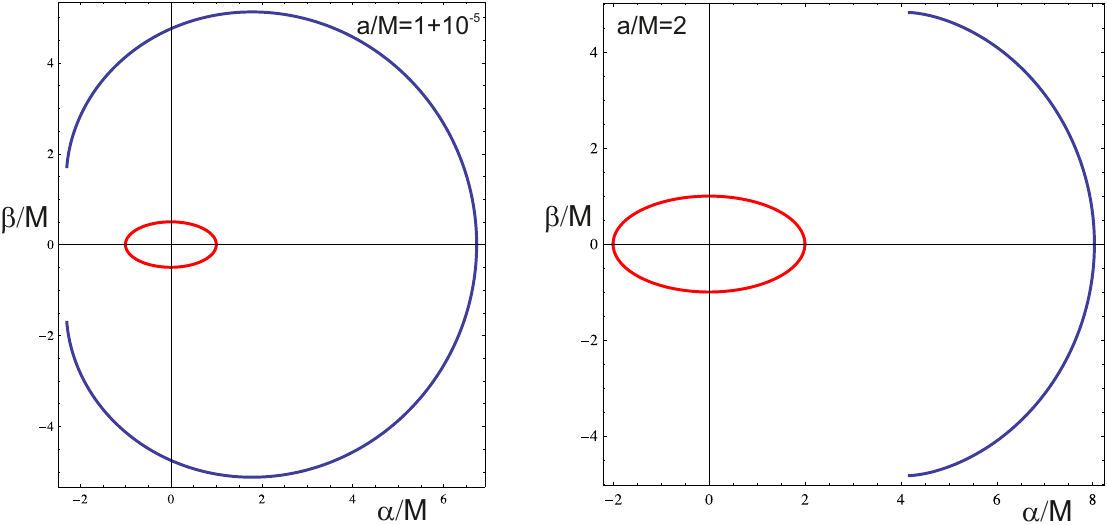}
\caption{\label{B&RSS} The silhouette of a topologically trivial SR-superspinar is characterized by an \emph{arc} (blue) (due to the existence of unstable photon orbits of constant $r\ (>0)$) and an ellipse (red) (due to the existence of a planckian region around the superspinar \emph{ring}). The parameters defining the BR superspinar have been chosen
$l/M=10^{-2}, L/M=1$ and $a/M=1+10^{-5}$ (left) $a/M=2$ (right). The inclination angle of observation in both cases is $\theta_i=\pi/3$. Note that the arcs are similar to the classical case if $l/M$ is small.}
\end{figure}

Examples of silhouettes of topologically trivial SR-superspinars are shown in figure \ref{B&RSS} with the same relationships $a/M$ as in the previous classical cases.

Note that non-topologically trivial SR-superspinars would also exhibit a dark spot, similar to Kerr superspinars. Thus, we demonstrate that the topology selected by Nature for SR-superspinars can, in principle, be tested through observations.

\subsection{Observing the ring}

As previously shown, it suffices to identify the ring of the superspinar in order to locate the planckian pseudotorus.
Only null geodesics with $\eta=0$ can reach the ring, but we have seen that they do so differently in Kerr superspinars and in SR-superspinars.

Let us first assume that the distant observer that wants to identify the ring is situated outside the equatorial plane of the superspinar. In this case, we have seen that the condition $\Theta\geq 0$ imposes on $\xi$ the condition $\xi^2\leq a^2$.

\begin{itemize}

\item For Kerr's superspinars only null geodesics on the equatorial plane can reach or exit the ring. Therefore, the observers outside the equatorial plane are not able to see the naked singularity at the ring.

\item For SR-superspinars the null geodesics reaching or leaving the ring (i.e., null geodesics with $\eta=0$) can be seen with celestial coordinates [see eqs.(\ref{celestial})]
     \[
\alpha=\frac{-\xi}{\sin\theta_i}\hspace{.5 cm}; \hspace{.5 cm} \beta= \pm \sqrt{(a-\xi)^2-\left(a \sin \theta_i-\frac{\xi}{\sin\theta_i} \right) ^2} .
\]
which satisfy
\[
\frac{\alpha^2}{a^2}+\frac{\beta^2}{a^2 \cos^2 \theta_i}=1.
\]
In this way, the ring is seen as an ellipse with semi-major axis $|a|$ and semi-minor axis $|a \cos \theta_i|$ by the observer. (Note that the null geodesics coming from this ellipse and reaching the observer have parameters $\xi$ between $\pm a \sin\theta_i$, which are precisely the values allowed for $\xi$ for geodesics with $\eta=0$ (subsec. \ref{eta0}) ). The consequences for the silhouette of an SR-superspinar are explicitly shown in Figure \ref{B&RSS}. The extreme cases occur when the observer is located along the $z$ axis ($\theta_i=0$), where the ring is seen as a circle, and when the observer is near the equatorial plane  ($\theta_i=\pi/2$), where the ring approximates a horizontal segment. Nevertheless, this limit requires a specific study (both in the classical and SR case) as follows.

\end{itemize}

\begin{figure}[htp]
\includegraphics[scale=.7]{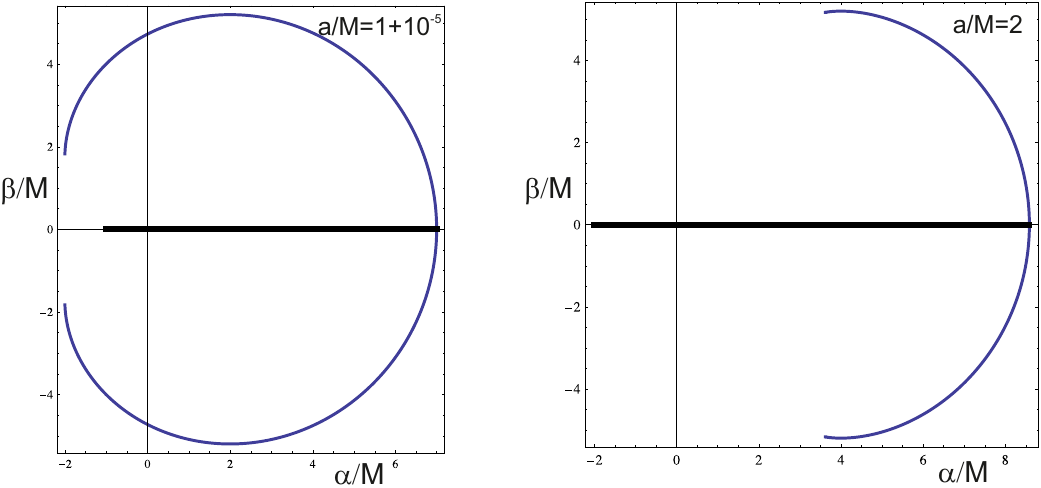}
\caption{\label{Kerreqplane} The silhouette of a Kerr superspinar as seen from the equatorial plane. The thick black line at $\alpha=0$ is caused by the ring singularity.}
\end{figure}

\begin{figure}[htp]
\includegraphics[scale=.7]{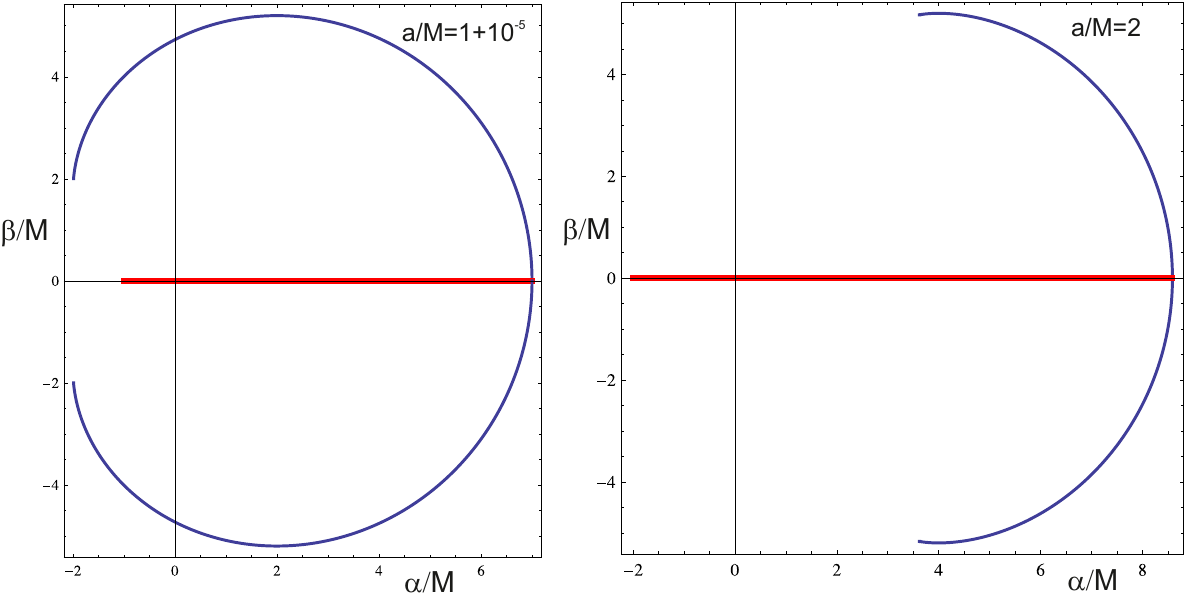}
\caption{\label{BReqplane} The silhouette of an SR-BR-superspinar as seen from the equatorial plane. The thick red line at $\alpha=0$ is caused by the null geodesics coming from the planckian region and reaching the distant observer.}
\end{figure}

Assuming the ring is viewed from the equatorial plane, now the condition $\Theta \geq 0$ does not place a restriction on $\xi$. However, the condition that the null geodesics are directed towards the future imposes (see eq. (\ref{dtdl})) that $\xi/a\leq 1$ (where $\xi=a$ is a special case in which we are dealing with a principal null geodesic).
There is another limit to $\xi$ for the geodesics that do reach the ring: If a null geodesic
moving towards smaller values of $r$ does not have a turning point with $r>0$ then necessarily will reach $r=0$. At the beginning of this section we have already treated the turning point conditions. In the case that concerns us now, to identify the turning point in celestial coordinates we have to use eqs.(\ref{celestial}) to find the $r_0^*$ such that $\beta(r_0^*)=0$, which provides the limit $\alpha_{max}\equiv \alpha(r_0^*)$.
In summary, assuming $a>0$, the ring as seen from the equatorial plane consists of a segment ($\beta=0$) between $-a\leq \alpha \leq \alpha_{max}$ (i.e., with right limit precisely on the arc). We show figures of superspinars as seen from the equatorial plane (both their arc and their ring) for Kerr superspinars and SR-superspinars in figures \ref{Kerreqplane} and \ref{BReqplane}, respectively.

\section{Stable circular orbits and possible instabilities}\label{SCOs}

Very recently it has been shown \cite{Filippo} (see also \cite{Cunha0}) that general axisymmetric horizonless UtraCompact Objects (UCOs) possessing unstable circular orbits must also possess at least one stable circular orbit for each rotation sense. The superspinars treated in this paper are a particular type of these UCOs for which we examined the unstable circular orbits in the previous section. Regarding their stable circular orbits, we have already shown that they must satisfy $R(r_0)=R'(r_0)=0$, which yielded equation (\ref{quadratic}) and that, if we require $R''(r_0)<0$, the only possible solution is $\xi_-$, what led us to $\eta_-$.

Restricting to those $r_0$ values for which $R''(r_0)<0$ determines the range of possible \textit{inner} stable orbits. They are inner because, under the assumptions of this paper, unstable orbits satisfy $r_0\geq r_0^*$ (with $R''(r_0^*)=0$), whereas stable orbits lie at $r_0< r_0^*$. Among these potential stable circular orbits, the physically valid solutions must also satisfy $\Theta \geq 0$ and $dt/d\lambda \geq 0$, as required by the null geodesic equations. Note also that if $r_0$ is such that $r_0-f_0 M_0=0$, then $R''(r_0)\rightarrow -\infty$, while $\xi_-$ and $\eta_-$ are also divergent.

\begin{figure}[htp]
\includegraphics[scale=.8]{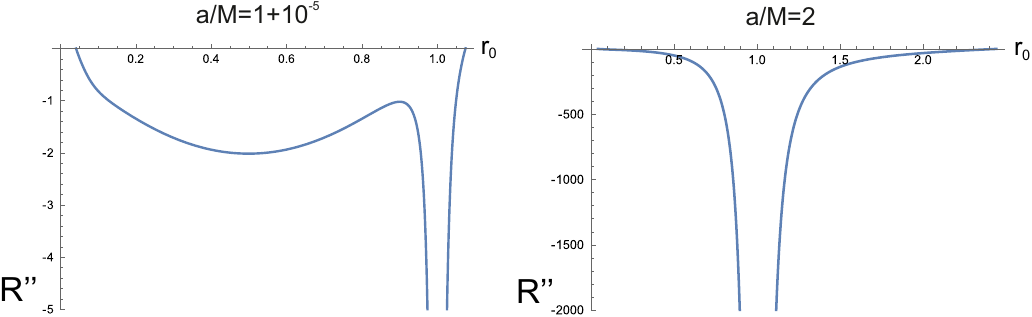}
\caption{\label{stableCO} The ranges of $r_0$ with $R''<0$ are exemplified in two BR superspinars with $l/M=10^{-2} $ and $L/M=1$.  To the left the model has $a/M=1+10^{-5}$ and $R''<0$ in the range $0.042<r_0<r0^*\simeq0.900$. To the right the model has $a/M=2$ and $R''<0$ in the range $0.041<r_0<r0^*\simeq2.442$. These models have a divergence for $R''$ at $r_0= 1.0003$.}
\end{figure}

Figure \ref{stableCO} illustrates the ranges of $r_0$ with negative values for $R''$ for two representative examples of SR-superspinars . Numerical analysis confirms that within these ranges, there are non-trivial intervals also satisfying $\Theta\geq 0$ and $dt/d\lambda\geq 0$, which correspond to physically admissible stable circular orbits.

The significance of stable circular orbits arises from the fact that their presence can indicate potential nonlinear instabilities, as perturbations may accumulate at these orbit locations
\cite{Keir,Cardoso,Cunha,Pani,Franzin}.
%\cite{Keir} \cite{Cardoso} \cite{Cunha} \cite{Pani} \cite{Franzin}.
Thus, SR-superspinars are \emph{potentially} unstable. Nevertheless, this issue remains unresolved. There is evidence supporting the stability of certain models \cite{Nakao,Roy}. Notably, a recent study \cite{Guo}, using fully nonlinear numerical evolution, demonstrates the existence of objects with stable circular orbits that remain stable.

\section{Summary}\label{conclu}

Recent advances have significantly improved our ability to test theoretical predictions related to high-curvature regions in the Universe. On one hand, direct observations of gravitational waves from astrophysical sources have been achieved by the LIGO-VIRGO-KAGRA collaborations \cite{ligo}. On the other hand, the Event Horizon Telescope (EHT) has provided us with images of black holes \cite{Aki}\cite{Aki2}. Additionally, substantial improvements are anticipated in the near future with the launch of the LISA project \cite{lisa} and the development of new ground-based observatories \cite{observ}. As a result, the study of physics in strong gravitational fields is becoming increasingly significant, not only in theoretical physics but also in astrophysical phenomenology.

Classical superspinars are often viewed as speculative and problematic objects as they have naked singularities, which would violate the cosmic censorship conjecture \cite{WCCC}\cite{SCCC}. Moreover, they also feature regions with effective negative mass (resulting in repulsive gravity in the $r < 0$ region) where causality violations are unavoidable. In contrast, we have shown that topologically trivial SR-superspinars do not violate the cosmic censorship conjecture, do not exhibit effective negative mass, and do not lead to causality violations, provided they are endowed with a non-negative mass function. This suggests that incorporating Quantum Gravity effects could render superspinars objects with plausible properties, warranting serious consideration of their existence. As we have seen, only the presence of stable circular orbits represents an unsolved problem for superspinars, as they may indicate potential nonlinear instabilities. This topic remains an active area of research. Current evidence suggests that at least some types of superspinars may remain stable \cite{Nakao,Roy,Guo}.

In this way, the phenomenology of SR-superspinars becomes an interesting subject of study. We have demonstrated that, in general, the silhouette of SR-superspinars will possess an arc. The arc in the silhouette should closely resemble the arc predicted in the silhouette of a Kerr superspinar if deviations from General Relativity are due to Quantum Gravity effects (and they are important at Planck's scale). The SR case will have slightly smaller arcs than its classical counterpart, but this effect would be practically impossible to observe.

On the other hand, a key difference between classical superspinars and topologically trivial SR superspinars is that the former are predicted to exhibit a visible dark spot due to the presence of the $r<0$ region beyond the disk, whereas the latter are expected to have silhouettes without dark spots.

Another significant difference between classical and SR superspinars is related to the observation of their rings. We have shown that the singular ring of classical superspinars would be practically impossible to observe in practice, as it would only be visible to distant observers positioned precisely in the equatorial plane of the superspinar. In contrast, SR-superspinars feature a Planckian region shaped like a pseudotorus, with the ring as part of its boundary. This region is observable by all distant observers, making it possible to study the Planckian region of superspinars and potentially gain extraordinary insights into Quantum Gravity.

\appendix

\section{Appendix: Differentiability at the disk}\label{apend}

In order to analyze the differentiability at the disk for SR-superspinars with metric (\ref{GGg}), let us proceed with an analysis similar to the one usually carried out for the classical rotating case.
Consider an observer crossing $r=0$ while moving along the $z$-axis ($x=y=0$). If we choose $r$ to be non-negative, then equation (\ref{defr}) implies that $r = |z|$ along the observer's trajectory, so that along this path
\begin{equation}\label{gttz}
g_{tt}=-1+ \frac{2 \mathcal M (|z|) |z|}{z^2+a^2}.
\end{equation}
The numerator in the fraction indicates that the derivative of this metric component along the axis, as well as the Christoffel symbols and the extrinsic curvature of the surface, can be discontinuous across the disk depending on the chosen mass function. A well-known instance of this discontinuity occurs when the mass function is constant: Kerr's solution. As previously mentioned, the differentiability issues in the Kerr solution can be resolved by analytically extending the spacetime through $r = 0$, allowing for negative values of $r$.

At first glance, the situation for general SR rotating objects appears more favorable. If the SR object has a mass function $\mathcal{M}(r) \sim r^n$ with $n \geq 3$ around $r = 0$, then the metric component along the trajectory (\ref{gttz}) will not have differentiability issues at $z = 0$ ($\partial_z g_{tt}(z=0) = 0$). In fact, it will be at least $C^n$,\footnote{Specifically, it will be $C^n$ if $n$ is even and $C^\infty$ if $n$ is odd.} suggesting that an extension through the disk may not be necessary for regular rotating objects.
To prove this conjecture, one must examine more than just a specific trajectory intersecting the disk and consider more than a single metric component. Notably, as we approach a point on the disk ($x^2 + y^2 < a^2$), according to equation (\ref{defr}), the function $r$ approaches zero whenever $z$ approaches zero, and vice versa.
If one chooses to avoid an extension with $r<0$ (i.e., if one chooses to work with a topologically trivial global spacetime) we get, solving for $r$ in (\ref{defr}), that around $z=0$\footnote{Note that we assumed $a>0$. (If not, here we should replace $a\rightarrow |a|$).}
\[
r \simeq \frac{a}{\sqrt{a^2-(x^2+y^2)}} |z|
\]
If we introduce this into the metric component (\ref{gttz}) and consider a mass function $\mathcal M (r) \sim r^n$ with $n\geq 3$ around $r=0$, we see that the metric component takes the form
\[
g_{tt} \simeq -1+\frac{f(x,y) |z|^{n+1}}{g(x,y) z^2+a^2},
\]
where $f$ and $g$ are finite differentiable functions in the disk. In this way, $g_{tt}$ is differentiable at the disk. (Specifically, again ($\partial_z g_{tt}(z=0)=0$) and, in fact, $g_{tt}$ is at least $C^n$ at the disk).
The reader can easily verify that a similar situation arises for the other metric components. However, it is important to note that the metric will not be analytic at the disk, regardless of the value of $n$. This is because not all metric components will be infinitely differentiable.
For example, even if the particular metric component (\ref{gttz}) for odd $n$ is $C^\infty$ in the disk, other metric components like
\[
g_{xz}=\frac{2\mathcal{ M}(r) r^2 z (a y+x r)}{(a^2+r^2)(a^2 z^2+r^4)}
\]
are not. ($g_{xz}$ is $C^n$ for odd $n$). Nevertheless, since usually the metric is required to be at least $C^2$ \cite{H&E}, a $C^n$ metric with $n\geq 3$ at the disk is more than enough.


\begin{thebibliography}{99}


\bibitem{stu2012}
Stuchlík Z and Schee J 2012 {\it Class. Quantum Grav.} {\bf 29} 65002

\bibitem{G&Hof}
Gimon EG and Hofava P 2009 {\it Physics Letters B} {\bf 672} 299 (arXiv:0706.2873 [hep-th])
% A favor de la existencia de superspinars como primordial remnants

\bibitem{NJGK2017}
Nakao K, Joshi PS, Guo J-Q, Kocherlakota P, Tagoshi H, Harada T, Patil M and Królak A 2018 {\it Phys. Lett. B} {\bf 780} 410

\bibitem{SHT}
Stuchlík Z, Hledík S and Truparová K 2011 {\it Class. Quantum Grav.} {\bf 28} 155017
%Superspinars inestables, pero alguna posibilidad de observación

\bibitem{BCH}
Bardeen JM, Carter B and Hawking SW 1973 {\it Commun. Math. Phys.} {\bf 31} 161

\bibitem{WCCC}
Penrose R 1969 {\it Riv. Nuovo Cim.} {\bf 1} 252 (2002 {\it Gen. Rel. Grav.} {\bf 34} 1141).
%La singularidad proveniente de un colapso debería estar oculta a observadores distantes

\bibitem{Li&Bambi}
Li Z and Bambi C. 2013 {\it Phys.Rev.D} {\bf 87} 124022 (arXiv:1304.6592 [gr-qc])
% Generar superspinars partiendo de un agujero negro

\bibitem{YZWL}
Yang S-J, Zhang Y-P, Wei S-W, Liu Y-X 2022 {\it JHEP} {\bf 04} 066 (arXiv:2201.03381 [gr-qc])
%Dado un Quantumm Black Hole se puede destruir su horizonte para dejar de ser un Black Hole

\bibitem{EH2023}
Eichhorn A and Held A 2023 {\it J. Cosmol. Astropart. Phys.} {\bf 01} 32
%Convierten BH en superspinar. Consideran que pueden venir geodésicas de r<0

\bibitem{H&Maeda}
Hioki K and Maeda K 2009 {\it Phys. Rev. D} {\bf 80} 024042 (arXiv:0904.3575 [astro-ph.HE])

\bibitem{B&F}
Bambi C and Freese K 2009 {\it Phys. Rev. D} {\bf 79} 43002	
%Suponen una región cuántica en el core que atrapa a los fotones en r>0
%They assume a ellipsoidal planckian region that traps all photons falling into it

\bibitem{Vries2000}
de Vries A 2000 {\it Class. Quantum Grav.} {\bf 17} 123
%Seguramente el primero en hablar de los dark spots (r<0) en Kerr-Newman, pero mal

\bibitem{SS2010}
Stuchlík Z and Schee J 2010 {\it Class. Quantum Grav.} {\bf 27}	215017	

\bibitem{TT24}
Tavlayan A, Tekin B 2024 {\it Class. Quantum Grav.} {\bf 41} 65004

\bibitem{KG21}
Kumar R, Ghosh SG 2021 {\it Class. Quantum Grav.} {\bf 38} 85010	

\bibitem{TorresInt}
Torres R 2023 {\it Phys. Rev. D} {\bf 108} 084008 (arXiv:2307.12096 [gr-qc])

\bibitem{IsraelThin}
Israel W 1966 {\it Nuovo Cimento B} {\bf 44} 1 ; 1967 {\it Nuovo Cimento B} {\bf 48} 463.

\bibitem{Israel}
Israel W 1970 {\it Phys. Rev. D} {\bf 2} 641

\bibitem{Hamity}
Hamity V 1976 {\it Phys. Lett. A} {\bf 56} 77

\bibitem{G&V}
Gibbons GW and Volkov MS 2017 {\it Phys. Rev. D} {\bf 96} 024053 (arxiv:1705.07787 [hep-th])

\bibitem{A-A}
Azreg-A\"{\i}nou M 2014 {\it Phys. Rev. D} {\bf 90} 064041 (arXiv:1405.2569 [gr-qc])

\bibitem{B&M}
Bambi C and Modesto L 2013 {\it Phys. Lett. B} {\bf 721} 329 (arXiv:1302.6075 [gr-qc])

\bibitem{FLMV}
Franzin E, Liberati S, Mazza J and Vellucci V 2022 arXiv:2207.08864 [gr-qc]

\bibitem{LGS}
De Lorenzo T, Giusti A and Speziale S 2016 {\it Gen. Rel. and Grav.} {\bf 48} 31. Corrigendum at {\it Gen. Rel. and Grav.} {\bf 48} 111

\bibitem{Maeda}
Maeda H 2022 {\it JHEP} {\bf 11}  108 (arXiv:2107.04791 [gr-qc])

\bibitem{MFL}
Mazza J, Franzin E and Liberati S 2021 {\it JCAP} {\bf 04} 082

\bibitem{eye}
Simpson A and Visser M 2022 {\it JCAP} {\bf 03} 011 (arXiv:2111.12329 [gr-qc])

\bibitem{D&G}
Dymnikova I and Galaktionov E 2015 {\it Class. Quant. Grav.} {\bf 32} 165015 (arXiv:1510.01353 [gr-qc])

\bibitem{Ghosh}
Ghosh S G 2015 {\it Eur. Phys. J. C} {\bf 75} 532 (arXiv:1408.5668 [gr-qc])

\bibitem{Tosh}
Toshmatov B, Ahmedov B, Abdujabbarov A and Stuchlik Z 2014 {\it Phys. Rev. D} {\bf 89} 104017 (arXiv:1404.6443 [gr-qc])

\bibitem{R&T}
Reuter M and Tuiran E 2011 {\it Phys. Rev. D} {\bf 83} 044041

\bibitem{TorresExt}
Torres R 2017 {\it Gen. Rel. and Grav.} {\bf 49} 74 (arXiv:1702.03567 [gr-qc])

\bibitem{BMR}
Bambi C, Modesto L and Rachwal L 2017 {\it JCAP} {\bf 05} 003 (arXiv:1611.00865 [gr-qc])

\bibitem{G&H}
Gomes H and Herczeg G 2014 {\it Class.Quant.Grav.} {\bf 31} 175014 (arXiv:1310.6095 [gr-qc])

\bibitem{Buri}
Burinskii A 2002 {\it Czech.J.Phys. } {\bf 52} C471

\bibitem{C&M}
Caravelli F and Modesto L 2010 {\it Class.Quant.Grav.} {\bf 27} 245022 (arXiv:1006.0232 [gr-qc])

\bibitem{S&S}
Smailagic A and Spallucci E 2010 {\it Phys. Lett. B} {\bf 688} 82 (arxiv:1003.3918 [hep-th])

\bibitem{GG}
G\"{u}rses M and G\"{u}rsey F 1975 {\it J. Math. Phys.} {\bf 16} 2385
%Lorentz Covariant Treatment of the Kerr-Schild Metric

\bibitem{TorresReg}
Torres R and Fayos F 2017 {\it Gen. Rel. and Grav.} {\bf 49} 2 	(arXiv:1611.03654 [gr-qc])

\bibitem{TorresCap}
Torres R 2023 {\it Regular Black Holes: Towards a New Paradigm of Gravitational Collapse} 421-446 Springer Nature, Singapore. ISBN: 978-981-9915-96-5 	(arXiv:2208.12713 [gr-qc])

\bibitem{B&R}
Bonanno A and Reuter M 2000 {\it Phys. Rev. D} {\bf 62} 043008 (arXiv:0002196 [hep-th])

\bibitem{Hay2006}
Hayward S A 2006 {\it Phys. Rev. Lett.} {\bf 96} 031103

\bibitem{H&E}
Hawking SW and Ellis GFR 1973 {\it The large scale structure of space-time} Cambridge University Press, Cambridge

\bibitem{Carter1968a}
Carter B 1968 {\it Phys. Rev.} {\bf 174} 1559

\bibitem{Tsuka}
Tsukamoto N 2018 {\it Phys. Rev. D} {\bf 97} 064021 (arXiv:1708.07427 [gr-qc])

\bibitem{AAAG}
Abdujabbarov A, Amir M, Ahmedov B and Ghosh SG 2016 {\it Phys. Rev. D} {\bf 93} 104004 (arXiv:1604.03809 [gr-qc])

\bibitem{ASG}
Ahmed F, Singh DV and Ghosh SG 2022 {\it Gen. Rel. and Grav.} {\bf 54} 21 	(arXiv:2002.12031 [gr-qc])

\bibitem{Am&G}
Amir M and Ghosh SG 2016 {\it Phys. Rev. D} {\bf 94} 024054 (arXiv:1603.06382 [gr-qc])

\bibitem{E&H}
Eichhorn A and Held A 2021 {\it JCAP} {\bf 05} 073 	(arXiv:2103.13163 [gr-qc])

\bibitem{LGPV}
Lamy F, Gourgoulhon E Paumard T sand Vincent FH  2018 {\it Class. Quantum Grav.} {\bf 35} 115009 (arXiv:1802.01635 [gr-qc])
%Muy citable: Consideran r<0 -de la extensión del RBH de Hayward-y obtienen dark spots

\bibitem{L&B}
Li Z and Bambi C 2014 {\it JCAP} {\bf 1401} 041 (arXiv:1309.1606 [gr-qc])

\bibitem{S&S0}
Stuchl\'{\i}k Z and Schee J 2019 {\it Eur. Phys, J} {\bf 79} 44

\bibitem{BCY}
Brahma S, Chen CY and Yeom DH 2021 {\it Phys. Rev. Lett.} {\bf 126} 181301 (arXiv:2012.08785 [gr-qc])

\bibitem{HGE}
Held A, Gold R and Eichhorn A 2019 {\it JCAP} {\bf 1906} 029 (arXiv:1904.07133 [gr-qc])

\bibitem{ligo}
See, for example, \url{https://en.wikipedia.org/wiki/List_of_gravitational_wave_observations} for a list of gravitational wave observations

\bibitem{Aki}
Akiyama K \textit{et al.} [Event Horizon Telescope] 2019 {\it Astrophys. J. Lett} {\bf 875} L1, L2, L3, L4, L5, L6

\bibitem{Aki2}
Akiyama K \textit{et al.} [Event Horizon Telescope] 2022 {\it Astrophys. J. Lett} {\bf 930} L12, L13, L14, L15, L16, L17

\bibitem{lisa}
Barausse E \textit{et al.} 2020 {\it Gen. Rel. and Grav.} {\bf 52} 81

\bibitem{observ}
Kalogera V \textit{et al.} arXiv:2111.06990 [gr-qc]

\bibitem{SCCC}
R. Penrose, Singularities of Spacetime, Theoretical Principles in Astrophysics and Relativity (A78-43851
19-90), Chicago University Press, Chicago (1978).
%Las soluciones de la RG deberian ser globalmente hiperbólicas

\bibitem{Filippo}
Di Filippo F 2024 {\it Phys. Rev. D} {\bf 110} 084026 (arXiv:2404.07357 [gr-qc])

\bibitem{Cunha0}
Cunha PVP, Berti E and Herdeiro C 2017 {\it Phys. Rev. Lett.} {\bf 119} 251102 (arXiv:1708.04211 [gr-qc])

\bibitem{Keir}
Keir J 2016 {\it Class. Quantum Grav.} {\bf 33} 135009 (arXiv:1404.7036 [gr-qc])

\bibitem{Cardoso}
Cardoso V, Crispino LCB, Macedo CFB, Okawa H and Pani P 2014 {\it Phys. Rev. D} {\bf 90} 044069 (arXiv:1406.5510 [gr-qc])

\bibitem{Cunha}
Cunha PVP, Herdeiro C, Radu E and Sanchis-Gual N 2023 {\it Phys. Rev. Lett.} {\bf 130} 061401 (arXiv:2207.13713 [gr-qc])

\bibitem{Pani}
Pani P, Barausse E, Berti E and Cardoso V 2010 {\it Phys.Rev.D} {\bf 82} 044009 (arXiv:1006.1863 [gr-qc])

\bibitem{Franzin}
Franzin E, Liberati S, Vellucci V 2024 {\it JCAP} {\bf 01} 020 (arXiv:2310.11990 [gr-qc])

\bibitem{Nakao}
Nakao K, Joshi PS, Guo JQ, Kocherlakota P, Tagoshi H, Harada T, Patil M, Krolak A 2018 {\it Phys. Let. B}
{\bf 780} 410 (arXiv:1707.07242 [gr-qc])

\bibitem{Roy}
Roy R, Kocherlakota P and Joshi PS (arXiv:1911.06169 [gr-qc])

\bibitem{Guo}
Guo G, Wang P and Zhang Y (arXiv:2403.02089 [gr-qc])

\end{thebibliography}
\end{document}